\documentstyle[12pt,amsmath,epsfig]{article}                                                      
                                                      
\newlength{\dinwidth}                                                      
\newlength{\dinmargin}                                                      
\def\lapproxeq{\lower .7ex\hbox{$\;\stackrel{\textstyle                                                      
<}{\sim}\;$}}                                                      
\def\gapproxeq{\lower .7ex\hbox{$\;\stackrel{\textstyle                                                      
>}{\sim}\;$}}

\def\bea{\begin{eqnarray}}                                                      
\def\eea{\end{eqnarray}}

\begin{document}                                                      
\titlepage    
\vspace*{1in}    
\begin{center}    
{\Large \bf Screening effects in the ultrahigh energy neutrino interactions}\\    
\vspace*{0.4in}    
Krzysztof Kutak and Jan   Kwieci\'nski   
\\    
\vspace*{0.5cm}    
  
{\it H. Niewodnicza\'nski Institute of Nuclear Physics,    
 Krak\'ow, Poland} \\    
\end{center}    
\vspace*{1cm}    
\centerline{(\today)}    
    
\vskip1cm    
\begin{abstract}
We study possible saturation effects in the total cross-sections describing interaction of  ultrahigh energy 
neutrinos with nucleons.  This analysis is performed within the two 
approaches, i.e. within  the Golec-Biernat W\"usthoff saturation model  and within 
the scheme unifying the DGLAP and BFKL dynamics incorporating non-linear 
screening effects which follow 
from the Balitzki-Kovchegov equation. The structure functions in both approaches are constrained 
by HERA 
data.   It is found that screening effects   
  affect extrapolation of the neutrino-nucleon total cross-sections to ultrahigh neutrino 
energies $E_{\nu}$  and 
reduce their magnitude by 
a factor equal to about 2  at $E_{\nu} \sim 10^{12}GeV$.  This reduction   
 becomes   
amplified by nuclear shadowing in the case     
of the neutrino-nucleus  cross-sections and approximate estimate  of this effect is 
performed.   
\end{abstract}
                                                   
\section{Introduction}
Ultrahigh energy neutrinos are one of the components of the spectrum of particles 
that reach Earth. As they interact weakly with matter their propagation through 
interstellar space is not attenuated. This is the reason why they are unique
carriers of information about distant objects such as GRB (Gamma Ray Bursts), 
AGN (Active Gallactic Nuclei) etc. which are most probably their sources.
This information can be studied by neutrino telescopes \cite{HALZEN}.\\
Attenuation of neutrinos traversing the Earth and their detection depend upon 
the cross-sections describing the interaction of neutrinos with nucleons and atomic nuclei. 
Ultrahigh energy neutrino interactions with nucleons are sensitive upon  the behaviour of the 
nucleon structure functions at extremely small values of the Bjorken parameter  
$x$ and relatively large scales $Q^2 \sim M_W^2$ \cite{GHANDI1,KMSNEUT,GRVNEUT,PREDAZZI}.  
Here, as usual $x=Q^2/(2pq)$, where $Q^2=-q^2$ with $p$ and $q$ denoting 
the four momentum of the nucleon and four momentum transfer between the leptons in the 
inelastic neutrino - nucleon interaction respectively.    The values of 
$x$ which can be probed can be several orders of magnitude smaller than those which are 
currently accessible at HERA \cite{HERA} and, for instance for  neutrino energies 
$E_{\nu} \sim 10^{12}GeV$ typical values of $x$ which contribute to the neutrino 
cross-sections can be as small as $x\sim 10^{-8}$.  Reliable estimate of the ultrahigh energy neutrino cross-sections 
does therefore require reliable extrapolation of the structure functions towards 
the  region of very small values of $x$ and large $Q^2$, i.e. far beyond the region 
which is currently accessible.    Existing estimates of the neutrino cross-sections with structure 
functions constrained by HERA data  are based on either DGLAP \cite{GHANDI1,GRVNEUT,PREDAZZI}
 or extended BFKL \cite{KMSNEUT}  
{\it linear} evolution which  neglects  non-linear  
 screening effects \cite{GLR,MQIU,SHDWG,MCLERR,KGBSHAD,BAL,KOV,MRSN}. Those effects are in general expected to slow down the increase of the 
parton distributions and of the cross sections with decreasing $x$ and  to   
reduce their magnitude.  Possible implications of screening effects for the estimate 
of the ultrahigh energy cross-sections have recently been discussed in ref.
\cite{NSAT1,NSAT2,BASU,MILIAN} with somewhat conflicting conclusions.  Thus in 
refs.\cite{NSAT1,NSAT2} it has 
been claimed  that the screening effects should play  negligible role in the estimate 
of the ultrahigh energy neutrino cross-sections due to the dominance of relatively 
large  scales $Q^2\sim M_W^2$.  On the contrary results obtained in ref. \cite{MILIAN} seem to imply 
that they may be significant and, moreover,when combined with 
the BFKL dynamics  can even  lead to enhancement of the cross-sections.\\

The purpose of this paper is to present relatively detailed and realistic 
estimate of the impact of the screening effects on the cross-sections 
desribing interaction of ultrahigh energy neutrino interactions.  
We perform this  analysis  within the two frameworks which 
incorporate screening effects, i.e. the Golec-Biernat-W\"usthoff (GBW) model \cite{GBW} and the unified 
BFKL/DGLAP scheme \cite{KMSNEUT,KMS} supplemented by the non-linear term in the 
corresponding evolution equations.   
This term will be  obtained  from the non-linear part of the 
Balitzki-Kovchegov (BK) equation \cite{BAL,KOV}.  In both cases the parton distributions and the resulting cross-sections 
will be constrained by the HERA data. The content of our paper is as follows. In the 
next Section we recollect basic
formulas describing the deep inelastic neutrino-nucleon scattering. In Section
3 we discuss description of deep inelastic scattering within the dipole picture 
and  present results for neutrino cross sections calculated within the GBW model.
Section 4 contains formulation of unified BFKL/DGLAP evolution equations  
supplemented by nonlinear screening effects and results for neutrino cross sections 
calculated within this approach.
Section 5 contains summary and conclusions.

\section{Basic formulas describing the deep inelastic neutrino scattering.}
The   deep inelastic neutrino scattering is illustrated by the diagram in Fig. 1. 
\begin{figure}[!h]
\centerline{\epsfig{file=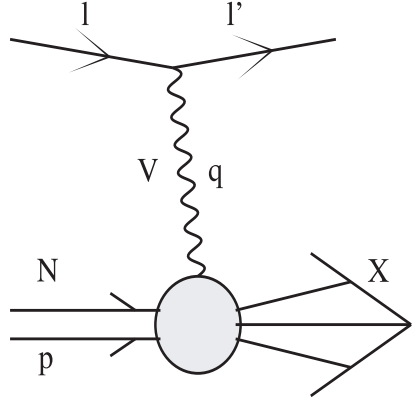,height=6cm,width=6cm}} 
\caption{Deep inelastic scattering.}
\label{fig:fig1}
\end{figure}  
It can proceed through  $W^{\pm}$ or $Z^0$ exchanges that corresponds  
to charged current (CC) or neutral current (NC) interactions respectively.  The charged current 
interactions correspond to the processes $\nu + N \rightarrow l^- + X$   
$( \bar \nu + N \rightarrow l^+ + X)$ with charged leptons $l^{\pm}$ in the final state while the 
neutral current 
interactions correspond to the processes $\nu + N \rightarrow \nu + X$   
$( \bar \nu + N \rightarrow \bar \nu + X)$.  The standard kinematical variables describing 
these processes are: 
$$
s=2ME
$$

$$
Q^2=-q^2
$$

$$
x={Q^2\over 2pq} 
$$

\begin{equation}
y={pq\over ME}
\label{kvar}
\end{equation}
where $M$ is the nucleon mass, $E$ denotes the neutrino energy while  $p$ and $q$ are the four 
momenta of the nucleon and of the exchanged boson respectively.  The cross-sections 
describing the deep inelastic neutrino scattering are expressed in the following way 
in terms of the structure functions $F_2^{CC,NC}(x,Q^2)$,  $F_L^{CC,NC}(x,Q^2)$ 
and $F_3^{CC,NC}(x,Q^2)$:   
$$
{\partial ^2 \sigma_{\nu,\bar \nu}^{CC,NC}\over \partial x\partial y}
={G_F^2ME\over \pi}\left({M_i^2\over Q^2+M_i^2}\right)^2
$$

\begin{equation}
\left[{1+(1-y)^2\over 2}F_2^{CC,NC}(x,Q^2)-{y^2\over 2}F_L^{CC,NC}(x,Q^2)
\pm y\left(1-{y\over 2}\right) xF_3^{CC,NC}(x,Q^2)\right]
\label{csx}
\end{equation}
where $G_F$ is the Fermi constant and $M_i$ denotes the mass of the charged 
($W^{\pm}$) or neutral ($Z^0$) gauge boson.\\

In the QCD improved parton model  
 the structure functions $F_{2,3}(x,Q^2)$ are expressed in terms of the 
(scale dependent) quark and antiquark distributions \cite{ROBERTS}.  
Thus for the isoscalar target $N={n+p\over 2}$ we have: 

\begin{equation}
F_2^{CC}(x,Q^2)=x[u_v(x,Q^2)+ d_v(x,Q^2)]+
2x[\bar u(x,Q^2) + \bar d(x,Q^2)+s(x,Q^2)+ c(x,Q^2)] 
\label{f2cc}
\end{equation}

\begin{equation}
F_3^{CC}(x,Q^2)=u_v(x,Q^2)+ d_v(x,Q^2)   
\label{f3cc}
\end{equation}

$$
F_2^{NC}(x,Q^2)={(L_u^2+L_d^2+R_u^2+R_d^2)\over 4} \times 
$$

\begin{equation}
\{x[u_v(x,Q^2)+ d_v(x,Q^2)]+
2x[\bar u(x,Q^2) + \bar d(x,Q^2)+s(x,Q^2)+ c(x,Q^2)]\}
\label{f2nc}
\end{equation}

\begin{equation}
F_3^{NC}(x,Q^2)={(L_u^2+L_d^2-R_u^2-R_d^2)\over 4} 
[u_v(x,Q^2)+ d_v(x,Q^2)]
 \label{f3nc}
\end{equation}

where the chiral couplings can be expressed  in terms of the Weinberg angle $\theta_W$
$$
L_u=1-{4\over 3} sin^2\theta_W
$$ 

$$
L_d=-1+{2\over 3} sin^2\theta_W
$$ 

$$
R_u=-{4\over 3} sin^2\theta_W
$$ 

\begin{equation}
R_d={2\over 3} sin^2\theta_W
\label{chcoupl}
\end{equation}

The quantities which are relevant for the quantitative description of the penetration of 
ultrahigh energy neutrinos through Earth and their detection are the neutrino cross-sections 
integrated over available phase space at the given neutrino energy. These integrated 
cross-section are given 
by: 
\begin{equation}
\sigma^{CC,NC}_{\nu,\bar \nu}(E)=\int _{Q_{min}^2}^s dQ^2\int_{Q^2/s}^1 dx {1\over xs}
{\partial ^2 \sigma_{\nu,\bar \nu}^{CC,NC}\over \partial x\partial y}
\label{sigint}
\end{equation}
with $y=Q^2/(xs)$. In equation (\ref{sigint}) we have introduced the minimal 
value $Q_{min}^2$ of $Q^2$ in order to stay in the deep inelastic region.  
In our calculations we set $Q_{min}^2=1GeV^2$.  
In the "low" energy region $s<M_i^2$ the integrated cross-sections rises linearly 
with $E$ and in this region interaction with valence quarks dominates.  
In the high energy region  the contribution of valence quarks saturates and the energy dependence 
of $\sigma^{CC,NC}_{\nu,\bar \nu}(E)$ is driven by the small $x$ behaviour of 
the sea quark distributions \cite{KMSNEUT}.  
It is this part of the cross-sections which will be analysed in our paper.\\

Existing numerical estimates of the ultrahigh energy cross-sections are based 
upon extrapolation of parton distributions towards the very small $x$ region 
using linear (DGLAP and/or BFKL)  QCD evolution equations 
\cite{GHANDI1,KMSNEUT,GRVNEUT}.  At small $x$ the dominant partons are the gluons 
and the sea quark distributions are driven by the gluons through the $g\rightarrow q \bar q$ 
transitions.  The linear QCD evolution generates indefinite increase of gluon distributions 
with decreasing $x$ that implies similar increase of the sea quark distributions and 
of the structure functions $F_2^{CC,NC}(x,Q^2)$ and $F_L^{CC,NC}(x,Q^2)$.  This increase 
is tamed  by the non-linear screening effects which lead to saturation 
\cite{GLR} - \cite{KOV}.  
Efficient way of introducing saturation can be realised using the colour dipole framework 
in which the DIS at low $x$ is viewed as the result of the interaction 
of the colour $q \bar q$ dipole which the gauge bosons fluctuate to as illustrated 
in Fig. 2.
\begin{figure}[!h]
\centerline{\epsfig{file=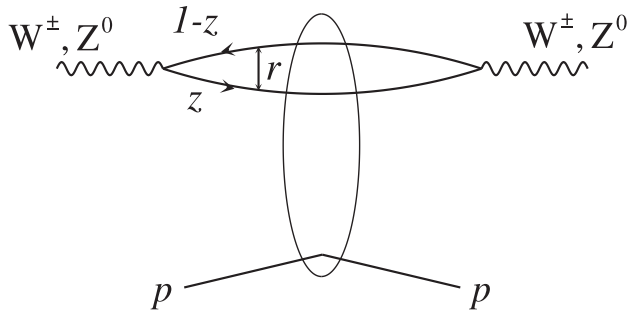,height=6cm,width=9.6cm}} 
\caption{Schematic representation of the dipole picture \cite{GBW}.}
\label{fig:fig2}
\end{figure} 
Very succesful semiphenomenolgical analysis of $ep$ DIS at low $x$ has been performed within this framework 
by Golec-Biernat and W\"usthoff \cite{GBW} and in the next Section we apply this model 
for the estimate  of the saturation effects in the ultrahigh energy 
neutrino cross-sections.          
\section{DIS in the dipole picture and ultrahigh energy neutrino interactions}
The DIS structure functions in the dipole picture can be written in the following form 
\cite{GBW}:  
\begin{equation}
F_{T,L}^{CC,NC}(x,Q^2)={Q^2\over 4 \pi^2} \int d^2{\bf r} \int_0^1 dz
|\bar \psi^{W,Z}_{T,L}(r,z,Q^2)|^2\sigma_d(r,x). 
\label{dcsx}
\end{equation}
In this equation $r$ denotes the transverse size of the $q \bar q$ dipole, $z$ the 
longitudinal momentum fraction carried by a quark and  $\bar \psi^{W,Z}_{T,L}(r,z,Q^2)$ 
are proportional to the wave functions of the (virtual) charged or neutral gauge bosons corresponding to their 
transverse or longitudinal polarisation 
($F_{T}^{CC,NC}(x,Q^2)=F_{2}^{CC,NC}(x,Q^2)-F_{L}^{CC,NC}(x,Q^2)$).  Explicit expressions 
for $\bar \psi^{W,Z}_{T,L}(r,z,Q^2)$ are given below. The cross-section 
$\sigma_d(r,x)$ describes interaction of the colour $q \bar q$ dipole with the nucleon.  
In the GBW model $\sigma_d(r,x)$ has the following form: 
\begin{equation}
\sigma_d(r,x)=\sigma_0\left[1-exp\left(-{r^2\over 4 R_0^2(x)}\right)\right]
\label{dipolecsx}
\end{equation}
The most crucial element in this model is the adoption of the $x$-dependent saturation 
radius $R_0(x)$ which scales the $q\bar q$ separation in the dipole cross-section.  Saturation radius 
is the decreasing function with decreasing  $x$ and is parametrised as below: 
\begin{equation}
R_0^2(x)={1\over Q_0^2} \left({x\over x_0}\right)^{\lambda}
\label{r02}
\end{equation}
with $Q_0^2=1GeV^2$.  
The three parameters of the model $\sigma_0,\lambda$ and $x_0$ 
were fitted to inclusive DIS data from HERA for $x<0.01$.  We shall use the 
 following values $\sigma_0=29.12 mb$, $\lambda=0.2777$ and $x_0=0.41\times 10^{-4}$ 
which were obtained from the fit with four flavours.\\

In the limit $r \rightarrow \infty$ we have $\sigma_d \rightarrow \sigma_0$ , 
i.e. the dipole  cross-section exhibits saturation property.  The fact that the dipole cross-section 
is limited by the energy independent cross-section can be regarded as the unitarity 
bound. In the limit $r\rightarrow 0$ the dipole cross-section vanishes reflecting the colour 
transparency.\\

The GBW model, which has proved to be phenomenologically very succesful 
in describing HERA data and has embodied saturation can be used for the estimate of the 
UHE neutrino cross-sections. In our calculation of UHE neutrino-nucleon cross-section we consider 
only four flavours ($u,d,s,c$).  The corresponding $q \bar q$ dipoles 
which contribute to Cabibbo favoured transitions are  $ u \bar d  (d \bar u)$,  
$ c \bar s (s \bar c)$ for charged currents and 
$ u \bar u, d \bar d, c \bar c, s \bar s$ for neutral currents respectively.  In our 
calculations we shall assume  massless quarks.  This approximation is reasonable for very high energy neutrinos.  
Possible contrtibution of heavy quarks $(b,t)$ where the 
mass parameters cannot be neglected is found to be relatively small 
\cite{KMSNEUT}. \\

The dipole model describes well the deep inelastic scattering at small $x$, but it becomes 
inaccurate at large and moderately small values of $x$.  This is closely linked with the fact that it 
neglects theoretical expectations concerning behaviour of quark distributions in the $x\rightarrow 1$ limit. 
This effect can be approximately taken into account by multiplying the structure functions 
by a factor  $(1-x)^{2n_s-1}$ which follows from the constitutent counting rule where $n_s$ denotes 
the number of spectator quarks.  Since the dipole model represents the sea quark contribution we set 
$n_s=4$.  \\

The functions $\bar \psi^{W,Z}_{T,L}(r,z,Q^2)$ corresponding to the sum over dipoles 
corresponding to massles quarks are given by the following formulas: 
\begin{equation}
|\bar \psi^{W}_{T}(r,z,Q^2)|^2={6\over \pi^2}[z^2+(1-z)^2]\bar Q^2K_1^2(\bar Q r)
\label{wt}
\end{equation}     
 
\begin{equation}     
|\bar \psi^{W}_{L}(r,z,Q^2)|^2={24\over \pi^2}z^2(1-z)^2 Q^2K_0^2(\bar Q r)
\label{wl}
\end{equation}     

\begin{equation}         
|\bar \psi^{Z}_{T}(r,z,Q^2)|^2={3\over 2 \pi^2}(L_u^2+L_d^2+R_u^2+R_d^2)[z^2+(1-z)^2]\bar Q^2K_1^2(\bar Q r)
\label{zt}
\end{equation}     
 
\begin{equation}     
|\bar \psi^{Z}_{L}(r,z,Q^2)|^2={6\over \pi^2}(L_u^2+L_d^2+R_u^2+R_d^2)z^2(1-z)^2 Q^2K_0^2(\bar Q r)
\label{zl}
\end{equation}  

where 
\begin{equation}
\bar Q^2=z(1-z)Q^2
\label{qbar2}
\end{equation}
and $K_{0,1}(u)$ are the Mc Donald's functions.\\ 

In Figures 3 and 4 we show results for the ultrahigh  neutrino cross-sections calculated within the GBW saturation model and 
confront them with the estimate based upon the unified BFKL/DGLAP framework which 
ignored saturation effects \cite{KMSNEUT}. We can see that the cross-sections calculated 
within the GBW model are at ultrahigh energies $E\sim 10^{12} GeV$ about a factor two 
smaller than those which were estimated within the scheme incorporating BFKL and DGLAP 
evolution without screening corrections.  
\newpage
\begin{figure}[!h]
\centerline{\epsfig{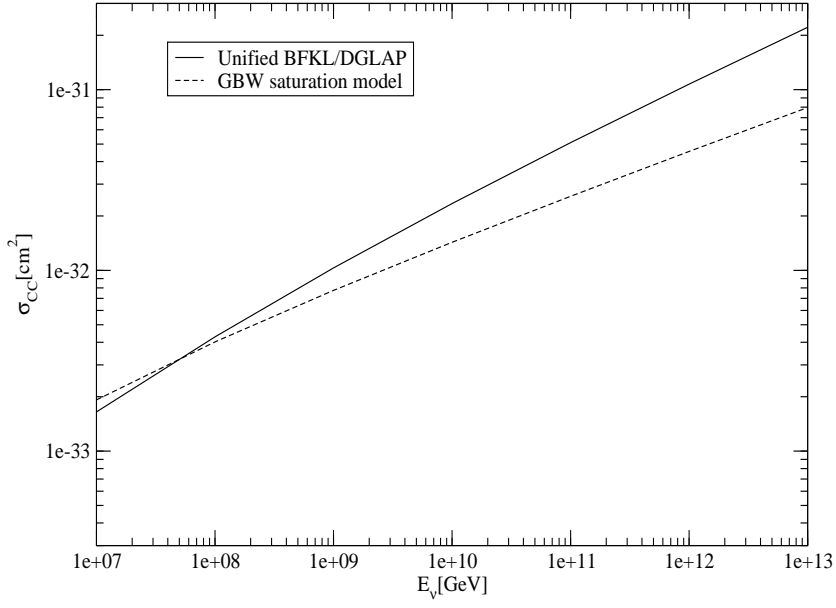}} 
\caption{The prediction for the neutrino nucleon CC cross section 
obtained from the GBW saturation model.  For comparison 
we also show  results based on the (linear) unified BFKL/DGLAP  evolution.  
}
\label{fig:fig3}
\end{figure} 
\vspace{2cm}
\begin{figure}[!h]
\centerline{\epsfig{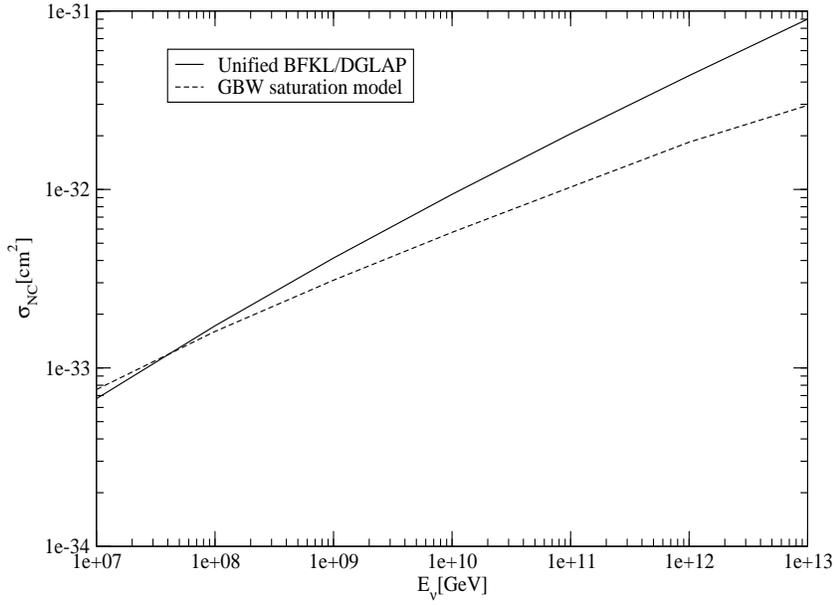}} 
\caption{As for Fig.~2, but for the NC interactions.}
\label{fig:fig4}
\end{figure}
\newpage

It should however be remembered that the GBW model has been tested 
phenomenologically for relatively low values of $Q^2$ and that it requires corrections 
incorporating DGLAP evolution \cite{BKGBK}.    It is therefore  necessary to perform an 
estimate of the neutrino cross-sections within the scheme that would include saturation effects 
together with complete QCD evolution as it will be described in the next Section.         
\section{Unified BFKL/DGLAP evolution with nonlinear screening effects and ultrahigh neutrino 
cross-sections}
We have shown in the previous Section that the ultrahigh energy neutrino cross-sections 
based upon the  GBW saturation model are at very high neutrino energies ($E>10^{12} GeV$ ) 
about a factor two  smaller than those calculated from the linear QCD evolution 
equations. It can however  be expected that  part of this reduction may just  be caused 
by the fact that the GBW model does not correctly include the QCD evolution effects \cite
{BKGBK}.  Let us recall that the dominant contribution to the 
neutrino cross sections comes from the region $Q^2 \sim M_W^2$ where the simple GBW model 
may not be sufficiently accurate.   It would therefore be desirable to discuss the ultrahigh energy cross-sections 
within the framework which contains both the QCD evolution effects and saturation.  
It is also of course very important that this framework should be based upon realistic 
parton distributions constrained by HERA data.  The framework should also contain all the QCD 
expectations concerning the small $x$ behaviour which follow from the BFKL dynamics as discussed, 
for instance, in ref. \cite{KMS}.  We shall therefore use the scheme developed 
in ref. \cite{KMS} containing the unified BFKL/DGLAP dynamics with subleading 
BFKL effects taken into account and supplement it by the screening contributions for the gluon 
distributions.   To be precise we shall include the non-linear screening term in the 
corresponding equation for the unintegrated gluon distribution $f(x,k^2)$, where $k^2$ is the transverse momentum squared of the 
gluon \cite{KMS}.  The structure functions 
are then calculated from the unintegrated gluon distributions using the $k_t$ 
factorisation prescription \cite{KTFAC}.  The extended system of the evolution equations with screening 
effects included then reads: 

$$
f(x,k^2)=\tilde f^{(0)}(x,k^2)
$$

$$
+ 2N_c{\alpha_s(k^2)\over 2 \pi} k^2\int_x^1{dz\over z} \int_{k_0^2}{dk^{\prime 2}\over 
k^{\prime 2}}\left \{{f\left({x\over z},k^{\prime 2}\right)\Theta\left({k^2\over z}-k^{\prime 2}
\right)-f\left({x\over z},k^2\right)\over |k^{\prime 2}-k^2|} +
{f\left({x\over z},k^2\right)\over [4k^{\prime 4}+k^4]^{{1\over 2}}}\right\}
$$

$$
+{\alpha_s(k^2)\over 2 \pi}\int_x^1{dz\over z}\left[\left(zP_{gg}(z)-2N_c\right)
\int_{k_0^2}^{k^2}{dk^{\prime 2}\over k^{\prime 2}}f\left({x\over z},k^{\prime 2}\right)+
zP_{gq}(z)\Sigma\left({x\over z},k^2\right)\right]
$$

\begin{equation}
-\left(1-k^2{d\over dk^2}\right)^2{k^2\over R^2}
\int_x^1{dz\over z}\left[\int_{k^2}^{\infty}
{dk^{\prime 2}\over k^{\prime 4}}\alpha_s(k^{\prime 2})ln\left(
{k^{\prime 2}\over k^2}\right)f(z,k^{\prime 2})\right]^2
\label{modf}
\end{equation}
The first three lines in equation (\ref{modf}) describe the linear unified BFKL/DGLAP 
evolution \cite{KMS}.  
Thus the second line of this equation corresponds to the BFKL evolution \cite{GLR,BFKLLO}.      
The constraint $\Theta\left({k^2\over z}-k^{\prime 2}\right)$ reflects the so called consistency 
constraint \cite{CCONST} which generates dominant part of the subleading 
BFKL corrections \cite{BFKLNLO,SALAM}. The two terms in the third line in eq. (\ref{modf}) correspond to the DGLAP 
effects generated by that part of the splitting function $P_{gg}(z)$ which is not 
singular in the limit $z\rightarrow 0$ and by the quarks respectively, with   $\Sigma(x,k^2)$ 
corresponding  to the singlet quark distributions
\begin{equation}
\Sigma(x,k^2)=\sum_{q=u,d,s}(q+\bar q) +c + \bar c 
=V(x,k^2) +S_{uds} + S_c
\label{singlet}
\end{equation}  
 where $V, S_{uds}$ and $S_c$ denote the valence, the light sea quark and the charmed 
quark distributions respectively. The inhomogeneous term $\tilde f^{(0)}(x,k^2)$ is defined in  terms of the input (integrated) gluon 
distribution: 
\begin{equation}
\tilde f^{(0)}(x,k^2)={\alpha_s(k^2)\over 2 \pi}\int_x^1 dz P_{gg}(z){x\over z} g\left({x\over z},k_0^2\right)
\label{f0}
\end{equation}

The nonlinear  screening contribution  is given by the last term in 
equation (\ref{modf}) where $R$ 
denotes the radius within which the gluons are expected to be concentrated. 
 The structure of this contribution  follows  
from the Balitzki-Kovchegov equation \cite{BAL,KOV} adapted to the unintegrated gluon 
distribution $f(x,k^2)$ \cite{KMK} and the 
details concerning the structure of this term are briefly discussed in the Appendix A.  
The sea quark distributions which describe the structure functions  are 
calculated from the unintegrated gluon distributions using $k_t$ factorisation.\\

Unlike the case of the leading $ln(1/x)$ approximation equation (\ref{modf}) 
 cannot be reduced to the evolution equation in $ln(1/x)$ that makes its numerical analysis 
rather cumbersome.  This complication comes  from  the fact that the subleading BFKL effects 
and the non-singular DGLAP contribution introduce nontrivial $z$ dependence 
of the kernel.     
We have however observed that at small $x$ linear version of equation (\ref{modf}) can be 
very well approximated by 
an (effective) 
evolution equation in $ln(1/x)$ with the boundary condition provided at some moderately small value 
of $x$ (i.e. $x=x_0 \sim 0.01$).  The latter is obtained from the solution of linear 
version of equation  
(\ref{modf}) in the region $x>x_0$.  
To be precise we have observed that at small 
$x$ we can make the following approximations: 
\begin{enumerate}
\item The  consistency constraint $\Theta(k^2/z-k^{\prime 2})$ of the  BFKL kernel 
responsible for the  subleading BFKL  
effects is replaced by the following 
effective ($z$ independent)  term 
\begin{equation}
\Theta(k^2/z-k^{\prime 2}) \rightarrow \Theta(k^2-k^{\prime 2}) +
\left({k^2\over k^{\prime 2}}\right)^{\omega_{eff}}\Theta(k^{\prime 2}-k^2)
\label{bfkleff}
\end{equation}
This replacement is motivated by the structure of the consistency constraint 
in the moment space, i.e. 
\begin{equation}
\omega \int_0^1{dz \over z}z^{\omega}\Theta(k^2/z-k^{\prime 2})=\Theta(k^2-k^{\prime 2}) +
\left({k^2\over k^{\prime 2}}\right)^{\omega}\Theta(k^{\prime 2}-k^2)
\label{ccmom}
\end{equation}
\item We make the following replacement: 
\begin{equation}
\int_x^1{dz\over z}[zP_{gg}(z)-2N_c]
f\left({x\over z},k^{\prime 2}\right) \rightarrow \bar P_{gg}(\omega=0) f(x,k^{\prime 2})
\label{dglapeff}
\end{equation}
where $\bar P_{gg}(\omega)$ is a moment function of $zP_{gg}(z)- 2N_c$, i.e.  
\begin{equation}
\bar P_{gg}(\omega)=\int_0^1{dz\over z}z^{\omega}[(zP_{gg}(z)-2N_c]
\label{pggmom}
\end{equation}
This approximation corresponds to keeping the leading term in the expansion 
of $\bar P_{gg}(\omega)$ around $\omega=0$ that is a standard approximation 
at       low $x$ \cite{LOWX}.  
\item We neglect the quark contribution in the right hand side of equation (\ref{modf}).  
\end{enumerate}

Using these approximations equation (\ref{modf}) can be rearranged into the 
following form: 
$${\partial f(x,k^2)\over \partial ln(1/x)}= 
K_L^{eff}\otimes f
$$

\begin{equation}
- \int_{k_0^2}^{\infty} 
 dk^{\prime \prime 2} K_S(k^2,k^{\prime \prime 2})\left(1-k^{\prime \prime 2}
{d\over d k^{\prime \prime 2}}\right)^2\left({k^{\prime \prime 2}\over R^2}
\right)
\left[\int_{k^{\prime \prime 2}}^{\infty}{dk^{\prime 2}\over k^{\prime 4}} ln\left({k^{\prime 2}\over 
k^{\prime \prime 2}}\right)\alpha_s(k^{\prime 2})f(x,k^{\prime  2})\right]^2
\label{evmod}
\end{equation}
where 

$$
K_L^{eff}\otimes f=
$$

$$
2N_c{\alpha_s(k^2)\over 2 \pi} \int_{k_0^2}^{\infty}dk^{\prime \prime 2}
\left\{\delta(k^2-k^{\prime \prime 2})+\Theta(k^2-k^{\prime \prime 2})\bar P_{gg}(0) 
{\alpha_s(k^{\prime \prime 2})\over 2 \pi} 
exp\left[\bar P_{gg}(0)(\xi(k^2)-
\xi(k^{\prime \prime 2})\right]\right\}
$$

\begin{equation}
  k^{\prime \prime 2} \int_{k_0^2}^{\infty}{dk^{\prime 2}\over 
k^{\prime 2}}\left \{{f(x,k^{\prime 2})\left[
\Theta(k^{\prime \prime 2}-k^{\prime 2})+\left({k^{\prime \prime 2}
\over k^{\prime 2}}\right)^{
\omega_{eff}}\Theta(k^{\prime 2}-k^{\prime \prime 2})\right]-
f(x,k^{\prime \prime 2})\over |k^{\prime 2}-k^{\prime \prime 2}|} +
{f(x,k^{\prime \prime 2})\over [4k^{\prime 4}+k^{\prime \prime 4}]^{{1\over 2}}}
\right\}
\label{kleff}
\end{equation}

\begin{equation}
K_S(k^2,k^{\prime \prime 2})=\delta(k^2-k^{\prime \prime 2})+
\Theta(k^2-k^{\prime \prime 2})\bar P_{gg}(0) 
{\alpha_s(k^{\prime \prime 2})\over 2 \pi} 
exp\left[\bar P_{gg}(0)(\xi(k^2)-\xi(k^{\prime \prime 2})\right]
\label{ks}
\end{equation}

 and 

\begin{equation}
\xi(k^2)=\int_{k_0^2}^{k^2} {dk^{\prime 2}\over k^{\prime 2}} 
{\alpha_s(k^{\prime 2})\over 2 \pi}
\label{xi}
\end{equation}
Following ref. \cite{KMS} we set $k_0^2=1GeV^2$. 
The method of the solution of equation (\ref{evmod}) is described in the 
Appendix B.\\

At first we solved the linear version of equation (\ref{evmod}) 
with the non-linear term neglected starting from the initial conditions 
at $x=10^{-2}$ obtained from the solution of the exact equation.  
The parameter $\omega_{eff}$ was then obtained by fitting the solution of the linear 
version of approximate 
equation (\ref{evmod}) to the solution of linear version of exact equation  (\ref{modf})  
of the unified BFKL/DGLAP framework.  
This procedure gives $\omega_{eff}=0.2$.  
It turns out that the solution of the linear version of the approximate equation  
(\ref{evmod}) reproduces the  solution of the  
linear version of equation (\ref{modf}) within  
3\% accuracy in the region $10^{-2}>x>10^{-8}$ and $2GeV^2<k^2<10^6GeV^2$).\\
We next solved the non-linear equation (\ref{evmod})  setting $R=4GeV^{-1}$.  
The quark distributions defining the  
  structure functions $F_{2,L}^{CC,NC}$ were calculated from the $k_t$ factorisation \cite{KMSNEUT,KMS} 
\begin{equation}
2xq(x,Q^2)=\int{dk^2\over k^2}\int_x^{a_q(k^2)} {dz\over z} 
S_q^{box}(z,k^2,Q^2)f\left(
{x\over z},k^2\right)
\label{ktfac}
\end{equation}
where the impact factors corresponding to the quark box contributions 
to gluon-boson fusion 
process are the same as those used in ref. \cite{KMS} (see also \cite{BE}), i.e.: 
$$
S_q^{box}(z,k^2,Q^2)={Q^2\over 4 \pi^2 k^2}\int_0^1 d\beta d^2\mbox{\boldmath 
{$\kappa^{\prime}$}}
\alpha_s\delta(z-z_0)
$$

\begin{equation}
\left\{[\beta^2+(1-\beta)^2]\left({\mbox{\boldmath {$\kappa$}}\over D_{1q}}-
{\mbox{\boldmath {$\kappa$}}-{\bf k}\over D_{2q}}
\right)^2+[m_q^2 +4Q^2\beta^2(1-\beta)^2
\left({1\over D_{1q}}-{1\over D_{2q}}
\right)^2\right\}
\label{impf}
\end{equation}
where $\mbox{\boldmath 
{$\kappa^{\prime}$}}
=\mbox{\boldmath {$\kappa$}}-(1-\beta){\bf k}$ and 

$$ 
D_{1q}=\kappa^2+\beta(1-\beta)Q^2 + m_q^2
$$

$$ 
D_{1q}=({\bf \kappa-k})^2+\beta(1-\beta)Q^2 + m_q^2
$$

\begin{equation}
z_0=\left[1+{\kappa^{\prime 2} +m_q^2\over \beta(1-\beta)Q^2}+{k^2\over Q^2}\right]^{-1}
\label{ddz}
\end{equation}
 To be  precise in the calculation of the (effective) quark distributions 
appearing in the charged current structure function we use the impact factors 
(\ref{impf}) corresponding to the 
massles quarks and the (charmed) quark mass  effects are  included  in the threshold 
factors: 
\begin{equation}
a_{c,s}(k^2)=\left(1+{k^2+m_{c}^2\over Q^2}\right)^{-1}
\label{acs}
\end{equation}
The $k_t$ factorisation 
formulas (\ref{ktfac},\ref{impf}) contain subleading $ln(1/x)$ effects coming from the exact kinematics of the 
gluon-boson fusion process \cite{BNP1}.  Complete NLO corrections to the impact 
factors are discussed in \cite{IMPFNLO}.    
In the  impact factors corresponding to the neutral currents we use 
equation (\ref{impf}) with $m_u=m_d=m_s=0$ and $m_c=1.4GeV$.      
  We also include non-perturbative 
contributions according to the prescription defined in ref. \cite{KMS}.     
The valence quark distributions were taken from ref. \cite{GRV95}. In Figures 5 and 6  we  show results 
of our calculation for $\sigma_{CC}\equiv\sigma^{CC}_{\nu}(E)$ and $\sigma_{NC}\equiv\sigma^{NC}_{\nu}(E)$ with and without 
screening corrections included and confront them  
 with our previous estimate based upon the GBW model. 
\vspace{1.5cm}
\begin{figure}[!h]
\centerline{\epsfig{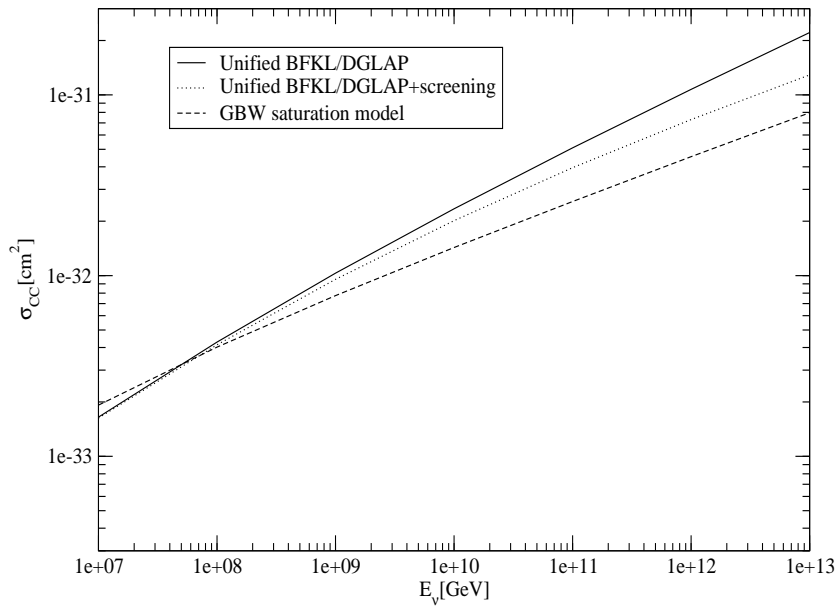}} 
\caption{The prediction for the neutrino nucleon CC cross section obtained from 
unified BFKL/DGLAP equation supplemented by screening effects.
For comparison we also present results based on the GBW saturation model and
the linear unified BFKL/DGLAP  evolutions}
\label{fig:fig5}
\end{figure} 
\begin{figure}[!h]
\centerline{\epsfig{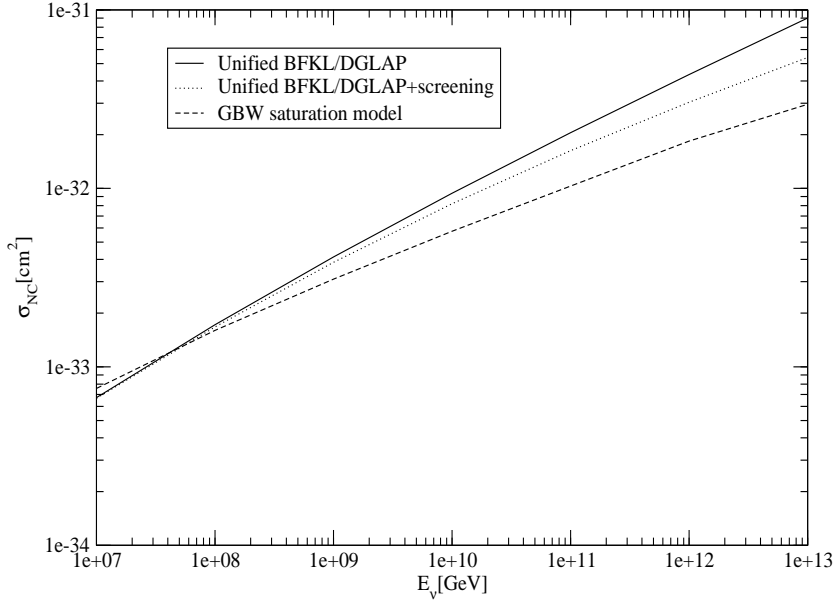}} 
\caption{As for Fig.~5, but for the NC interactions.}
\label{fig:fig6}
\end{figure}
 We can see that at ultrahigh energies  the cross-sections  
calculated within the unified BFKL/DGLAP framework supplemented by screening effects 
are bigger than those calculated from the simple GBW model.  
 The resulting cross-sections are still appreciably smaller than the cross-sections 
calculated within the linear BFKL/DGLAP framework with the screening effects neglected.\\

Reduction of the magnitude of the neutrino cross-section  is the consequence of 
the fact that the non-linear screening effects slow down increase of the structure 
functions with decreasing $x$.
In Fig. 7 we show the charged current structure function 
$F_2^{CC}(x,Q^2)$ plotted as the function of $x$ at $Q^2=M_W^2$ with and without 
screening corrections included. We can see that the screening effects reduce 
the magnitude of  $F_2^{CC}(x,Q^2=M_W^2)$ at $x=10^{-8}$ by almost a factor 
equal to two.\\
The screening effects in structure functions are generated through $k_t$ factorisation 
by the screening effects in the unintegrated gluon distribution $f(x,k^2)$ which satisfies 
the non-linear equation (\ref{evmod}).  The non-linear screening corrections generate 
the critical line $Q_c^2(x)$ which increases with decreasing $x$ which divides the $k^2,x$ 
plane into the two regions.   
In the region 
 $k^2<Q_c^2(x)$ the unintegrated gluon distribution saturates, i.e. 
$f(k^2,x) \sim R^2k^2h(x,k^2)$, where $h(x,k^2)$ is a slowly varying function of $x$ and $k^2$. 
In this region the unintegrated gluon distribution becomes then much smaller than 
the solution $f_{l}(k^2,x)$ of the linear version of equation (\ref{evmod}) 
which behaves approximately as $f_{l}(k^2,x) \sim x^{-\lambda}$ with 
$\lambda \sim 0.3$ over the entire region of $k^2$ \cite{KMSNEUT,KMS}.  In the region 
$k^2>Q_c^2(x)$ the non-linear screening contribution in the right hand side 
of equation (\ref{evmod})  becomes less important than the linear term and 
can be neglected for $k^2>>Q_c^2(x)$.  The magnitude 
of the unintegrated distribution continues  to be significantly  smaller than $f_l(x,k^2)$ over 
substantial range of $k^2$.  To be precise we have $f(k^2,x) \sim f(Q_c^2(x),x)$ 
with $f(Q_c^2(x),x)<f_l(k^2,x)$ in the 
region 
\begin{figure}[!h]
\centerline{\epsfig{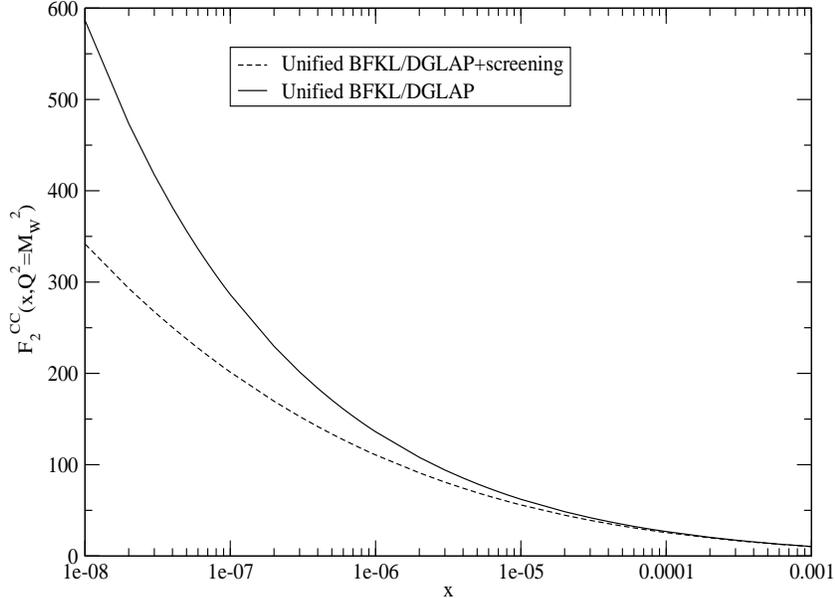}} 
\caption{The $F_{2}^{CC}(x,Q^2)$structure function obtained from the unified 
BFKL/DGLAP equation supplemented
by screening efects compared to results based on  the linear BFKL/DGLAP evolution.  
The function $F_{2}^{CC}(x,Q^2)$ is plotted as the function of 
$x$ for $Q^2=M_W^2$. }
\label{fig:fig7}
\end{figure}
\begin{equation}
\tilde Q^2(x)>k^2>Q_c^2(x)
\label{vgs}
\end{equation}
 where 
$\tilde Q^2(x)=Q_c^4(x)/\Lambda_{QCD}^2$ \cite{KSGEOM}. \footnote{Condition (\ref{vgs}) has   
a  simple origin.  It comes from the fact that  possible scaling violations in the region 
$k^2>Q_c^2(x)$      which modify boundary condition provided along the 
critical line $Q_c^2(x)$ are approximately controlled by the 'evolution length' 
$\tilde \xi(k^2,x)\sim\alpha_s(Q_c^2(x))
ln(k^2/Q_c^2(x))$.  Condition (\ref{vgs}) is  equivalent to the requirement 
$\tilde \xi(k^2,x)<<1$} It should be noted that $\tilde Q^2(x)>>Q_c^2(x)$.  
The screening effects do therefore  significantly reduce 
the corresponding contribution to the $k_t$ factorisation integrals (\ref{ktfac}) coming from
the region $k_0^2<k^2<\tilde Q^2(x)$. The integral over this region gives of course part of 
the leading twist contribution to the structure functions $F_2^{CC,NC}(x,Q^2)$ that does not 
vanish at large $Q^2$.  
This result that the screening effects at the structure function $F_2$  are  
appreciable  
even at such a large value of $Q^2 \sim M_W^2$  comes therefore from the fact that the screening effects 
contribute to the leading twist part of $F_2^{CC,NC}(x,Q^2)$.\\
The fact that the screening effects at $F_2^{CC,NC}(x,Q^2)$ can be important  
at $Q^2 \sim M_W^2$ and very small $x$ ($x \sim 10^{-8}$) implies that they may in turn have   
non-negligible influence on the ultrahigh energy neutrino cross-sections.  
It is this fact which makes our results significantly different from those presented in 
refs. \cite{NSAT2,BASU} where the saturation effects were confined to the modification 
of the structure functions  in the saturation region $Q^2 <Q_c^2(x)$ only.  The corresponding 
contribution to the UHE neutrino cross-section coming from the   integral 
over this region in equation (8)  is  very small and so  modifications 
of the structure functions in the saturation region alone have  negligible impact on the 
UHE cross-sections \cite{NSAT2,BASU} .

The fact that the cross-sections are sensitive upon the behaviour at very small $x$ 
and large scales $Q^2 \sim M_W^2$ implies that the effects which 
are formally subleading in $ln(1/x)$ but can  significantly affect both the $ln(1/x)$ and $Q^2$ evolutions
cannot be neglected. 
We illustrate this point in Fig. 8 where 
we show $F_2^{CC}(x,Q^2)$ for $Q^2=M_W^2$ calculated within three approximations: the leading $ln(1/x)$ 
BFKL framework, the extended BFKL framework which includes the subleading $ln(1/x)$ 
effects generated by the consistency constraint and the unified  BFKL/DGLAP 
scheme which includes besides the BFKL dynamics with subleading effects also 
the complete DGLAP evolution. 
The latter two frameworks contain effects which are subleading 
at  low $x$. 
\vspace{1cm}
\begin{figure}[!h]
\centerline{\epsfig{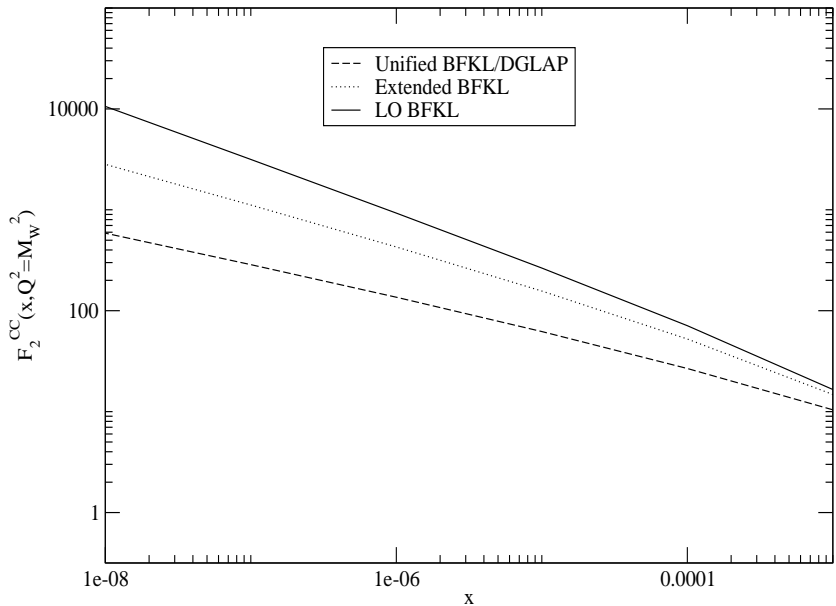}} 
\caption{The comparison of the $F_2^{CC}$
structure function calculated for $Q^2=M_W^2$ obtained from unified BFKL/DGLAP 
evolution, extended BFKL evolution and  
LO BFKL evolution.  Extended BFKL evolution corresponds to the BFKL equation 
with subleading effects generated by the consistency constraint but without the 
DGLAP effects.}
\label{fig:fig8}
\end{figure} 
For simplicity of presentation the non-linear screening effects are neglected in all 
three cases. We can see that both subleading $ln(1/x)$ effects play very important role 
and significantly reduce the magnitude of the structure function at large scale 
$Q^2 \sim M_W^2$ and very small 
values  of $x \sim 10^{-8}$.
Including these effects together with screening 
contribution is therefore important for getting a reliable extrapolation of the structure 
functions into the region of very small values of $x$ and large scales.\\  

Discussion of the cross-sections performed so far concerned screening effects on a nucleon 
target.  In the case of the neutrino-nucleus inelastic scattering 
further reduction of the magnitude of the total neutrino cross-sections due to nuclear 
shadowing is expected \cite{PARENTE}.  
In order to perform an indicative estimate of the possible nuclear shadowing 
effects for different values of the atomic numbers $A$ we just modify the strength 
of the non-linear of the non-linear term in eq. (\ref{modf})  by a factor $A^{1/3}$.  In Fig. 9 we show our 
results for the normalised neutrino-nucleus cross sections for different values 
of the atomic number $A$ varying from $A=12$ to $A=207$.
For comparison we show results for the neutrino-nucleon cross-section with and without screening effects.  
We see from this Figure that the nuclear shadowing can lead to 
further reduction of the cross-section.  
\vspace{1.3cm}
\begin{figure}[!h]
\centerline{\epsfig{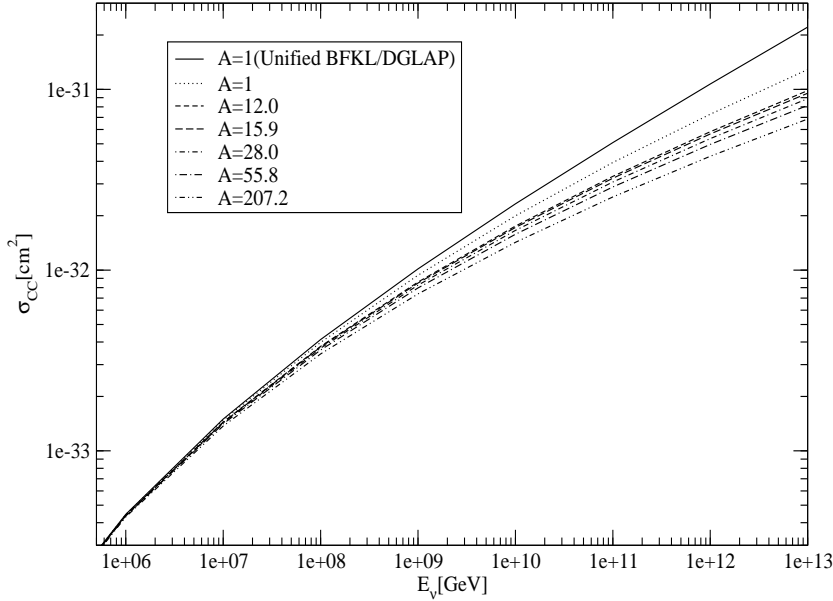}} 
\caption{The prediction for the neutrino nucleus CC cross section obtained from 
unified BFKL/DGLAP equation suplemented by screening effects. 
Cross section is calculated
for diffrent atomic numbers and normalised to nucleon. For comparison we also 
present 
results for neutrino nucleon CC cross section based on the (linear) 
unified BFKL/DGLAP  evolution.}
\label{fig:fig9}
\end{figure} 

\section{Summary and conclusions.} 
In this paper we have performed analysis of possible implications of the screening 
effects on the extrapolation of the neutrino-nucleon cross sections towards the 
ultrahigh  energy region.  Behaviour of the cross-sections in this region probes the 
structure functions at very small values of $x$ and relatively large scales 
$Q^2 \sim M_{W,Z}^2$.  The values of $x$ which can be probed can be as small as 
$10^{-8}$ and   it may  be expected that parton densities in this ultra small $x$ region  
should be affected by non-linear screening effects which tame the indefinite 
increase  of parton distributions generated by linear (BFKL and/or DGLAP) QCD 
evolution.  At first we have performed an estimate of the total neutrino-nucleon 
cross sections within the Golec Biernat - W\"usthoff saturatuion model.  In this 
model the  deep inelastic lepton scattering is viewed as the result of the 
interaction of the colour $q \bar q$ dipoles which the gauge boson fluctuates to.  
Important ingredient of the model is the fact that it incorporates saturation 
property of the total dipole-nucleon proton cross-section at large transverse 
separations between constituents of the dipole.  We have found that the neutrino 
total cross-sections obtained within this model are at ultrahigh 
neutrino energies  significantly smaller than those estimated from the linear 
BFKL/DGLAP evolution which neglects screening effects.  We have observed however that 
part of this reduction might have been caused by the fact that the GBW model did not 
include the QCD evolution effects and so it 
was not sufficiently accurate at very small values of $x$ and large value of 
the scale $Q^2 \sim M_{W,Z}^2$.  In order to overcome this potential deficiency of the GBW 
model 
we have performed an estimate of the cross section within the more elaborate framework 
based on the unified BFKL/DGLAP scheme supplemented by the non-linear screening effects.  
Contrary to the simple GBW model this framework contained complete BFKL and DGLAP evolution 
including the subleading BFKL contributions.  We have shown  that all these effects 
are important in the region of very small values of $x$ and large scales which is 
relevant for the interactions of ultrahigh energy neutrinos.  The non-linear screening 
effects were still found to  reduce appreciably the neutrino cross-sections at 
ultrahigh energies yet their effect turn out to be milder than in the case of 
the simple GBW model. We have also presented a very approximate estimate of the nuclear 
shadowing effects on further reduction of the cross-sections.\\

To summarise we have shown that the screening effects may play non-negligible 
role in the extrapolation of the  neutrino cross-sections towards the ultrahigh 
energies. 
\section*{Acknowledgments}
We thank Krzysztof Golec-Biernat, Alan Matin and Anna Sta\'sto for useful discussions.  
This research was partially supported by the Polish State Committee for Scientific Research 
(KBN) grants n0. 2P03B 05119, 5P03B 14420. 
                              
\section*{Appendix A}
In this Appendix we derive the non-linear shadowing term in the right hand side 
of eq. (\ref{evmod}) starting from the Balitzki-Kovchegov equation.  The basic quantity 
within this framework is the number of the colour dipoles $N(r,{\bf b},x)$ 
in a nucleon where 
$r$ denotes the transverse size of the dipole and $b$ is the impact parameter.  
The quantity  $N({\bf r},{\bf b},x)$ is closely related to the  total 
cross-section $\sigma(r,x)$ describing interaction of the $q \bar q$ colour dipole of 
transverse size $r$ with a nucleon
\begin{equation}
\sigma(r,x)=2\int d^2{\bf b}N({\bf r},{\bf b},x) 
\label{sivsn}
\end{equation}
The dipole cross-section is related to the unintegrated gluon distribution 
$f(x,k^2)$ \cite{BNP2}:
\begin{equation}
\sigma(r,x)={8 \pi^2 \over N_c}\int{dk\over k^3}[1-J_0(kr)]\alpha_s f(x,k^2)
\label{sigvsf}
\end{equation}
For simplicity we regard  $\alpha_s$ as fixed parameter and will put its argument at 
the end.    
It is convenient to introduce  the functions  
$\tilde N({\bf l},{\bf b},x)$ and $\tilde n(l,x)$ defined as below: 
\begin{equation}
  \tilde N({\bf l},{\bf b},x)=\int{d^2{\bf r}\over 2\pi  r^2}exp[i{\bf l r}]N({\bf r},{\bf b},x)
\label{tilden}
\end{equation}
\begin{equation}     
\tilde n(l,x)=\int  d^2 {\bf b}\tilde N({\bf l},{\bf b},x)
\label{tnlx}
\end{equation} 
 From equations (\ref{sivsn},\ref{sigvsf},\ref{tilden}, \ref{tnlx}) we get: 
\begin{equation}
\tilde n(l,x)={\pi^2 \over N_c}\int_{l^2}^{\infty}{dk^2\over k^4}\ln({k^2\over l^2})
\alpha_s f(x,k^2)
\label{nvsf}
\end{equation}
where we have used the following relation:
\begin{equation}
\int_0^{\infty}{dr\over r}J_0(lr)[1-J_0(kr)]=\Theta(k^2-l^2)ln({k\over l})
\label{lnkl}
\end{equation}
Equation (\ref{nvsf}) implies the following local relation between 
$\tilde n(l,x)$ and $f(x,l^2)$: 
\begin{equation}
(1-l^2{d\over dl^2})^2 l^2n(l,x)={\alpha_s\pi^2\over N_c}f(x,l^2)
\label{nvsfl}
\end{equation} 
In the large $N_c$ limit the function $N({\bf r},{\bf b},x)$ satisfies 
the Balitzki-Kovchegov equation \cite{BAL,KOV}: 
$$
N({\bf r_{01}},{\bf b},x)=N_0({\bf r_{01}},{\bf b},x)
$$

$$
+{\alpha_s N_c\over 2 \pi}\int_x^1{dz\over z}\{-2ln{r_{01}^2\over \rho^2}
N({\bf r_{01}},{\bf b},z)
$$

\begin{equation}
\int_{\rho}^{\infty} {d^2{\bf r_{2}}\over \pi}{r^2_{01}\over r^2_{02}r^2_{12}}\left[
2N\left({\bf r_{02}},{\bf b} +{1\over 2}{\bf r_{12}},z\right)-
N\left({\bf r_{02}},{\bf b} +{1\over 2}{\bf r_{12}},z\right)N\left({\bf r_{12}},{\bf b} -{1\over 2}{\bf r_{20}},z\right)
\right]\}
\label{kovr}
\end{equation}

The term linear in $N$ in the right hand side of equation (\ref{kovr}) corresponds to the 
right hand side of the BFKL equation (in transverse coordinate space)  
in the leading $ln(1/x)$ approximation and the 
nonlinear term describes  screening effects.  Taking the Fourier-Bessel transform of both 
sides of eq.(\ref{kovr}), integrating over ${\bf d^2 b}$ using an approximation 
$b>>1/2r_{20}$ and $b>>1/2r_{10}$ in the nonlinear term and assuming the following 
factorisation:
\begin{equation}
 \tilde N^2({\bf l}, {\bf b}, z)=\tilde n(l,x)S(b)
\label{tnvsn}
\end{equation}
with
\begin{equation}
\int d^2{\bf b} S(b)=1
\label{norm}
\end{equation}
we get
\begin{equation}
l^2\tilde n(l,x)=l^2 \tilde n^{0}(l^2,x)+{N_c \alpha_s\over \pi}\int_x^1{dz \over z}
\left[K\otimes l^2\tilde n - {1\over \pi R^2}l^2\tilde n^2(l,z)\right]
\label{kovl}
\end{equation}
where 
\begin{equation}
{1\over \pi R^2}=\int {\bf d^2b} S^2(b) 
\label{radius2}
\end{equation}
The kernel $K$ in equation (\ref{kovl}) is the LO BFKL kernel. Using equations 
(\ref{nvsf}) and (\ref{nvsfl}) we transform equation (\ref{kovl}) 
into an equation for the 
unintegrated gluon distribution: 

$$  
f(x,k^2)=f^{0}(x,k^2)+ 
$$
\begin{equation}
\int_x^1{dz \over z}\left\{{N_c \alpha_s\over \pi}K\otimes f 
-\left(1-k^2{d\over dk^2}\right)^2 \left({k^2\over R^2}\right)\left[\int_{k^2}^{\infty}
{dk^{\prime 2} \over k^{\prime 4}}\alpha_s(k^{\prime 2}) 
ln\left({k^{\prime 2}\over k^2}\right)f(z,k^{\prime 2})\right]^2\right\}
\label{kovf}
\end{equation}
where following ref. \cite{BKGBK} we set argument of $\alpha_s$  equal to $k^{\prime 2}$ 
in the non-linear term.  Finally supplementing the linear evolution by the subleading BFKL 
effects generated by the consistency constraint and the DGLAP 
contributions resulting from 
the non-singular parts of the $P_{gg}$ splitting 
function and quarks we get equation (\ref{modf}).  Following ref. \cite{KMS} we set argument 
of $\alpha_s$ in the linear term equal to $k^2$.    

\section*{Appendix B}
In order to solve equation (\ref{evmod}) we use the Tchebyshev interpolation 
formula: 
\begin{equation}
f(x,k^2)={2\over N}\sum_{n,i=0}^{N-1}v_nT_n(\tau_i)T_n(\tau_{k^2})f(x,k^2_i)
\label{czeb1}
\end{equation}
where $T_n(z)$ are the Tchebyshev polynomials, $v_0=1/2$ and $v_n=1$ for $n>0$.  
The variables $\tau_{k^2}$, $\tau_i$ and $k_i^2$ are defined as 
\begin{equation}
\tau_{k^2}={ln\left({k^2\over k_{max}k_{0}}\right)\over ln\left({k_{max}\over k_0}\right)}
\label{tauk2}
\end{equation}

\begin{equation}
\tau_i=cos\left({2i+1\over 2N}\pi\right)
 \label{taui}
\end{equation}

\begin{equation}
k_i^2=k_{max}k_{0}\left({k_{max}\over k_{0}}\right)^{\tau_i}
\label{ki2}
\end{equation}
where we set $k_0^2=1GeV^2$, $k_{max}^2=10^6 GeV^2$. Combining the Tchebyshev interpolation 
formula (\ref{czeb1}) with equation (\ref{evmod}) we reduce this equation to a system 
of the non-linear differential equations for the functions $f(x,k_i^2)$.  This system 
is solved 
using the standard Runge-Kutta method and the function $f(x,k^2)$  calculated 
from (\ref{czeb1}) for arbitrary value $k^2$ in the region 
$k_0^2<k^2<k_{max}^2$.  In the region $k^2>k_{max}^2$ which gives negligible 
contribution anyway we approximate the function $f(x,k^2)$ by $f(x,k_{max}^2)$.   

\end{document}